\documentclass[prl,reprint,flushbottom,showpacs]{revtex4-1}
\usepackage{epsfig}
\usepackage{amsmath}
\usepackage{amssymb}
\usepackage{amsthm} 
\usepackage{bm}
\usepackage{color}

\newcommand{\ket}[1]{\left|#1\right>}
\newcommand{\bra}[1]{\left<#1\right|}
\newcommand{\braket}[2]{\left< #1 \vphantom{#2} \right|\left. #2 \vphantom{#1} \right>}
\newcommand{\nn}{\nonumber\\}

\newcommand{\bea}{\begin{eqnarray}}
\newcommand{\ea}{\end{eqnarray}}
\newcommand{\eea}{\end{eqnarray}}
\newcommand{\ord}{{\cal O}}
\newcommand{\sumint}[1]

\begin{document}

\title{Revealing single-trap condensate fragmentation 
by measuring density-density correlations 
after time of flight
} 
\author{Myung-Kyun Kang and Uwe R. Fischer}

\affiliation{Seoul National University, Department of Physics and Astronomy \\  Center for Theoretical Physics, 
151-747 Seoul, Korea}

\begin{abstract}
We consider  ultracold bosonic atoms in a single trap in the Thomas-Fermi regime, forming many-body states 
corresponding to stable macroscopically fragmented two-mode condensates.
It is demonstrated that upon free expansion of the gas, the spatial dependence 
of the density-density correlations at late times provides a unique signature of fragmentation. 
This hallmark of fragmented condensate many-body states in  a single trap is due to the fact that 
time of flight modifies 
the correlation signal 
such that two opposite 
points in the expanding cloud become uncorrelated, {in distinction to a nonfragmented Bose-Einstein condensate, where they remain correlated}. 
\end{abstract}


\maketitle

{\em Introduction.}
The textbook definition of Bose-Einstein condensation 
consists in the existence of exactly one $\ord (N)$ (i.e., macroscopic) 
eigenvalue of the single-particle density matrix (SPDM) 
\cite{Penrose,Leggett,Pethick}, where $N\gg 1$ is the total number of particles.  
When interactions become sufficiently strong, the condensate is 
depleted by scattering processes \cite{Bogoliubov,Pethick}. 
A fundamental question then arises: Upon increasing
the interaction beyond a certain threshold, do fragmented condensates with two or more $\ord (N)$ eigenvalues of the SPDM exist \cite{Mueller}, or does the system cross over directly 
from a single condensate to nonmacroscopic fragments?

The phenomenon of fragmentation is well known when the externally applied potential provides deep 
double wells \cite{Spekkens}, or for the periodic extension deep optical lattices \cite{Greiner}, where the 
fragmented phase bears the name Mott insulator.
However, there has been the prevalent belief 
that for experiments performed with ultracold atoms of one given species 
in {\em single} (e.g., harmonic) traps a nonfragmented Bose-Einstein condensate is obtained, 
despite these experiments usually being conducted in the 
Thomas-Fermi (TF) limit, for which the kinetic energy is small compared to trapping and interaction energies.
That is, macroscopic condensate fragmentation is supposed in these experiments 
to not occur before three-body recombination \cite{Fedichev} 
destroys the condensate rapidly.  

On the other hand, recent work has demonstrated 
that condensate fragmentation is a genuine many-body phenomenon, and is intrinsically not describable within a simple mean-field theory (within an effective Gross-Pitaevskii theory) \cite{Bader,Fischer,Streltsov,Alon}. 
In a single trap, 
fragmentation 
occurs for repulsive interactions in the ground state \cite{Bader}, and for experimentally accessible TF parameters \cite{Fischer,Streltsov},   
against the expectation that for repulsive interactions no fragmentation is obtained \cite{Nozieres}. 
{In the TF limit, interaction thus can lead to the 
population of several macroscopically occupied orbitals.}
The (quasi-)continuity  of distribution amplitudes in Fock space has been shown 
to be responsible for the stability of fragmentation, 
also against thermal fluctuations \cite{FischerII}. 
This is in strong contrast with the unstable fragmentation occurring, 
e.g., in spin-orbit coupled gases \cite{Gopalakrishnan} or spinor gases \cite{Dan}, for which 
fragmented states are (superpositions of) exact Fock states \cite{Jackson}, i.e. have sharply peaked distributions in Fock space.    

An outstanding open question 
concerns the {\em detection} of fragmentation in a single trap, that is to verify conclusively that it indeed has taken place. {Fragmentation in the superfluid-Mott transition on optical lattices  is detected by the decrease of the visibility of the structure factor peaks \cite{Greiner}. This first-order correlation function measure of coherence, directly related to the SPDM in position space, $\rho_1({\bm r},{\bm r}')$ \cite{EsslingerJModOpt}, 
is in a single trap not operative.} This is primarily because in general the macroscopically occupied natural orbitals 
{(for a definition see below)}  
will significantly overlap 
in a potentially complicated fashion,  
in distinction to the multiple-well scenario, 
where they are well separated \cite{Spekkens,Greiner}. Unequivocally assigning fragmentation 
to the measured signal will thus be severely hampered.  
This difficulty becomes particularly relevant when the degree of 
fragmentation is relatively small. 

Detecting density-density correlations is by now a standard tool to discriminate one many-body phase from the other \cite{Altman}; the 
correlations  can be measured both {\em in situ} \cite{Hung}, and {\em ex situ}, that is after time of flight (TOF), cf., e.g., \cite{Foelling}. 
Motivated by this fact, we propose a readily implemented 
experimental procedure to determine whether a given condensate has fragmented.  
It is demonstrated that density-density correlations after TOF give a clear and unequivocal signature for the fragmentation. 
As we will show, counterintuitively, the essentially noninteracting expansion, which necessarily diminishes the density, magnifies the characteristic signature of fragmentation. 


{We first introduce some terminology. Expanding the field operator as  
$\hat \psi({\bm r})=\sum_i \psi_i(\bold{r})\hat{a}_i$, and writing the SPDM in its eigenbasis, 
{$\rho_1 ( {\bm r},{\bm r}' ) = \sum_i \lambda_i \psi^*_i({\bm r})\psi_i({\bm r}')$}, 
the corresponding orbitals $\psi_i({\bm r})$ are called {\em natural}.   
We then have $\big<\hat{a}^{\dagger}_i\hat{a}_j\big>=0,\, \forall\, i\neq j$, and the eigenvalue $\lambda_i=\big<\hat{a}^{\dagger}_i\hat{a}_i\big>$ is the occupation number of the natural orbital $\psi_i(\bold{r})$. A many-body state with more than 
one $\lambda_i=\ord(N)$ is a fragmented condensate.
We perform the calculation below 
for two macroscopically occupied orbitals, assuming that the thermal portion of atoms is negligible. 
The SPDM is then a (truncated) $2\times 2$ matrix, and the {\em degree of fragmentation} is defined by ${\mathcal F} =1-|\lambda_0-\lambda_1|/N$.
When both eigenvalues are $\ord(N)$, 
$\mathcal F$ is finite, and becomes maximal (unity) when they are both equal to $N/2$. }
Considering 
two macroscopic fragments is partly motivated by the recent study \cite{Streltsov}, finding a stepwise increase 
of the number of fragments from the single condensate 
upon increasing the interaction coupling.

For two orbitals (modes), the Fock space many-body state reads
\bea
\ket{\Psi}=\sum_{l=0}^{N}C_l\,\frac{(\hat{a}^{\dagger}_0)^{N-l}(\hat{a}^{\dagger}_1)^l}{\sqrt{(N-l)!l!}}\ket{0}
\equiv \sum_{l=0}^{N}C_l\,\ket{N-l,l}.
\ea
We assume the rather generic condition on the many-body amplitudes $C_l$, see Ref.\,\cite{Bader}, 
that they have a sharply peaked continuum limit distribution for the moduli, e.g., the Gaussian  
$|C(l)|=\left(\pi a^2\right)^{-1/4}\textrm{exp}[-\left(l-N/2-\mathcal{S}\right)^2/(2a^2)].$   
Here, the width of the distribution $a\propto \sqrt N$ and the shift  
$\mathcal{S}$ are 
given in terms of the parameters of a two-mode Hamiltonian in the trap, e.g.,  of the form 
$\hat H =\epsilon_0\hat a_0^\dagger\hat a_0
+\epsilon_1\hat a_1^\dagger\hat a_1 
+\frac{A_1}2 \hat a^\dagger_0\hat a^\dagger_0 \hat a_0 \hat a_0 
+\frac{A_2}2 \hat a^\dagger_1\hat a^\dagger_1\hat a_1\hat a_1
+(\frac{A_3}{2}\hat a_0^\dagger\hat a^\dagger_0
\hat a_1\hat a_1+ {\rm h.c.})  
+\frac{A_4}2 \hat a_1^\dagger \hat a_1 
\hat a_0^\dagger\hat a_0$, where $\epsilon_i$ are single-particle energies
and $A_i$ interaction couplings depending on the orbitals and the two-body interaction. We then have a maximum at $l_0$ ($=N/2+{\mathcal S}$ for the Gaussian distribution), whose relative width becomes very small when $N\gg 1$.
{Note that there are no single-particle tunneling terms $-\frac12 \Omega  
\hat a_0^\dagger \hat a_1 + {\rm h.c.}$ and number-weighed tunneling 
terms $\propto \hat n_0 \hat a_0^\dagger \hat a_1 + {\rm h.c.}$ or $\propto\hat n_1 \hat a_0^\dagger \hat a_1 + {\rm h.c.}$ when the two modes have even (0) and odd (1) parity, respectively (also see below).}
{We set 
the pair-exchange coupling $A_3>0$ (which is naturally of the same order as the other $A_i$ in a single trap  \cite{Bader,Fischer}). 
From energy minimization and  
the discrete time-independent Schr\"odinger equation $EC_l = \frac12 {A_3}(d_lC_{l+2}+d_{l-2}C_{l-2}) 
+ [\epsilon_0(N-l)+\epsilon_1 l+\frac12A_1(N-l)(N-l-1)
+\frac12A_2 l(l-1)+\frac12A_4 (N-l)l]C_l$, connecting $l$ ``sites'' in Fock space differing by 2, 
we obtain  $\textrm{sgn}(C_l C_{l+2})=-1$ ($C_l\in \mathbb{R}$ \cite{Lee}).
This entails a fragmented condensate many-body state 
due to the consequent condition 
sgn($C_l C_{l+1})=\pm (-1)^l$ \cite{Bader,Fischer}}. 


\begin{center}
\begin{figure}[b]
 \includegraphics[width=\columnwidth]{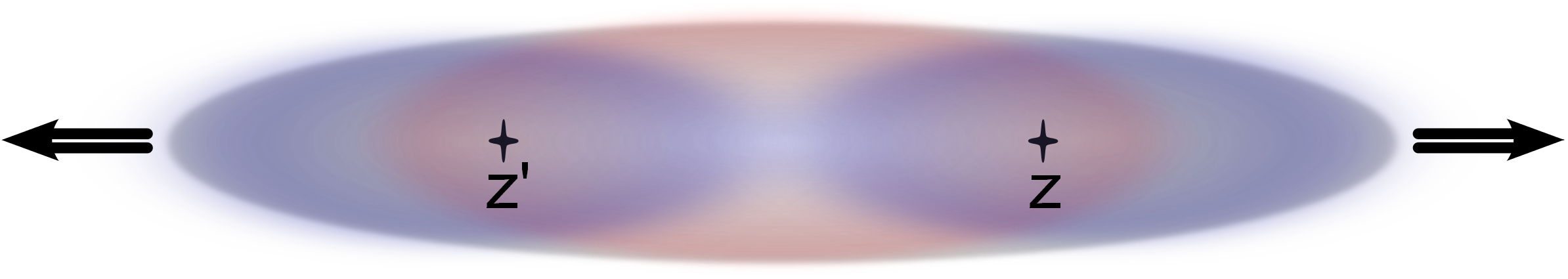} 
 \caption{\label{illus} Schematic of an axially freely expanding 
 quasi-1D gas 
 in a fragmented condensate many-body state.  
 The two macroscopically occupied orbitals are indicated by red and blue shaded areas.
 Density correlations are measured at two (opposite) points $z,z'$ in the 
 cloud at some given instant $t$.}
 \end{figure}
 \vspace*{-3em}
\end{center}

{\em Density-density correlations.} 
We focus from now on quasi-one-dimensional (quasi-1D) condensates, for which the largest degrees 
of fragmentation can be expected \cite{Fischer}. We also assume that the condensate is deep in the 
TF regime of large particle numbers \cite{Dunjko}.
The density expectation value in terms of the axial coordinate $z$, in the natural basis, reads $\rho(z)=\big<\hat{\psi}^{\dagger}(z)\hat{\psi}(z)\big>=
N_0 |\psi_0(z)|^2 +N_1|\psi_1(z)|^2$, where $N_i = \lambda_i = \big<\hat{a}^{\dagger}_{i}\hat{a}_{i}\big>$. 
The density-density correlation function (the the two-particle density matrix (TPDM) in position space \cite{remarkTPDM}) then takes the form 
\begin{multline} \label{twomodeddcorr}
\rho_2(z,z')=
\big<\hat{\psi}^{\dagger}(z)\hat{\psi}^{\dagger}(z')\hat{\psi}(z')\hat{\psi}(z)\big> \\
=
|\psi_0(z)|^2|\psi_0(z')|^2\big<\hat{a}^{\dagger}_{0}\hat{a}^{\dagger}_{0}\hat{a}_{0}\hat{a}_{0}\big>+ 0\rightarrow 1
\\
+\left(
|\psi_0(z)|^2|\psi_1(z')|^2+ 0\leftrightarrow 1 
\right)
\big<\hat{a}^{\dagger}_{0}\hat{a}^{\dagger}_{1}\hat{a}_{1}\hat{a}_{0}\big> \\
+2\Re\left[\psi^*_0(z)\psi^*_1(z')\psi_0(z')\psi_1(z)\big<\hat{a}^{\dagger}_{0}\hat{a}^{\dagger}_{1}\hat{a}_{1}\hat{a}_{0}\big>\right.\\
\left. +\psi^*_0(z)\psi^*_0(z')\psi_1(z')\psi_1(z)\big<\hat{a}^{\dagger}_{0}\hat{a}^{\dagger}_{0}\hat{a}_{1}\hat{a}_{1}\big>\right].
\end{multline}
It is for given orbitals $\psi_i(z)$ 
prescribed by the TPDM elements  $\big<\hat{a}^{\dagger}_{i}\hat{a}^{\dagger}_{j}\hat{a}_{k}\hat{a}_{l}\big>$,  
which are in turn determined by the many-body amplitudes $C_l$. 
The last line contains the pair-exchange term, which  
decides whether the many-body ground state in a single trap is fragmented \cite{Bader}.

For simplicity, the initial orbitals are assumed to fulfill 
that  $\psi_0(z,0)$ is an even real function of $z$ with $\psi_0(z)=\psi_0(-z,0)\in\mathbb{R}$, 
$\psi_1(z,0)$ is an odd real function of $z$ with $\psi_1(z,0)=-\psi_1(-z,0)\in\mathbb{R}$, i.e.
have definite parity in the trap \cite{expect}. 
We define $w$ as a (finite) common width measure of the orbitals, 
which {is, e.g., a variational parameter determined by the competition 
of interaction and trapping \cite{Fischer}.} 
In what follows, $w=1$ is used as the unit of length, as well as $\hbar =m=1$, with $m$ the boson mass. 

Calculating the TPDM elements from the continuum limit for $C_l$, we have to $\ord(1/N)$
\cite{NPCremark} 
\bea
\big<\hat{a}^{\dagger}_{0}\hat{a}^{\dagger}_{0}\hat{a}_{0}\hat{a}_{0}\big>={N}_0^2,~   \big<\hat{a}^{\dagger}_{1}\hat{a}^{\dagger}_{1}\hat{a}_{1}\hat{a}_{1}\big>={N}_1^2,~ \nn  \big<\hat{a}^{\dagger}_{0}\hat{a}^{\dagger}_{1}\hat{a}_{1}\hat{a}_{0}\big>={N}_0{N}_1,~ \big<\hat{a}^{\dagger}_{0}\hat{a}^{\dagger}_{0}\hat{a}_{1}\hat{a}_{1}\big>=-{N}_0{N}_1 . \label{TPDMcont}
\ea
This result remains valid as long as the $C_l$ distribution is centered at $l_0\sim \ord (N)$ with a width $\ll N.$

{Turning off the trap potential in the weakly confining axial direction only \cite{microtrap}, 
cf.\,Fig.\,\ref{illus}, after a short initial period of rapid expansion, for $t\gg 1$, 
the gas will expand ballistically \cite{ballnote}.
One can then apply the noninteracting propagator 
to the initial orbitals}
\bea
\psi_j(z,t) 
=\sqrt{\frac{1}{2\pi i w^2_t}} \exp \left[\frac{iz^2}{2w^2_t}\right] \tilde{\psi}_j(z,t),\qquad
 w_t = \sqrt{t}, \label{freepropagator}
\ea
where $\tilde{\psi}_j(z,t)= \exp [\frac{-iz^2}{2w^2_t}]
\,\int dz'\psi_j(z',0)\exp[\frac{i(z-z')^2}{2w^2_t}]$. 
{At late times, $\tilde{\psi}_j(z,t)$ has the meaning of a Fourier transform with respect to the variable pair $(z',z/w_t^2)$ to first order in $z'/w_t$, $\psi_j(z',0)$ remaining spatially confined.} 

\begin{center}
\begin{figure}[t]
\hspace{-1.9em}
\includegraphics[width=0.85\columnwidth]{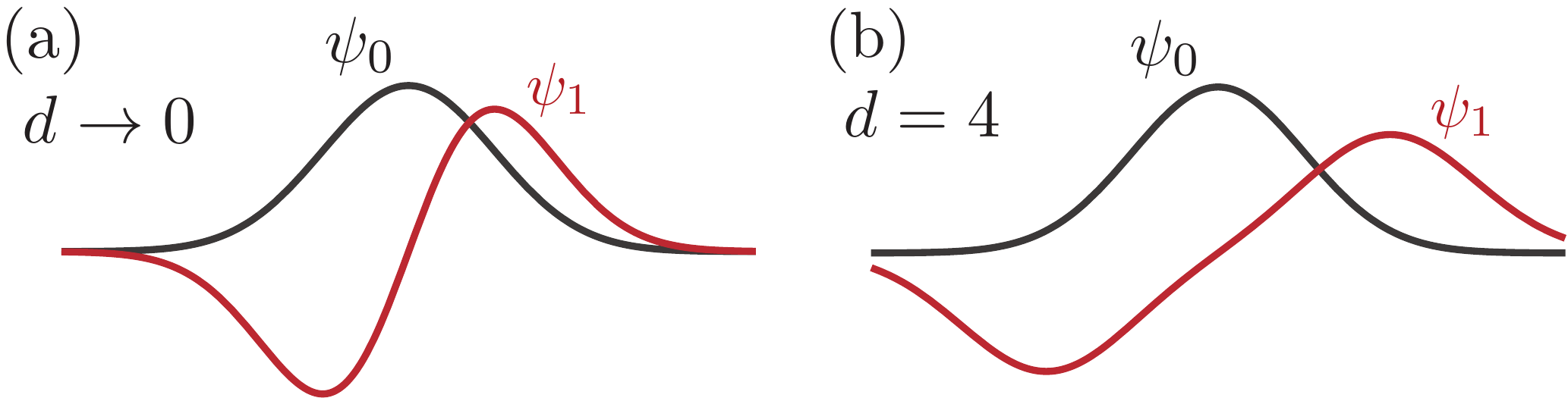}
\includegraphics[width=\columnwidth]{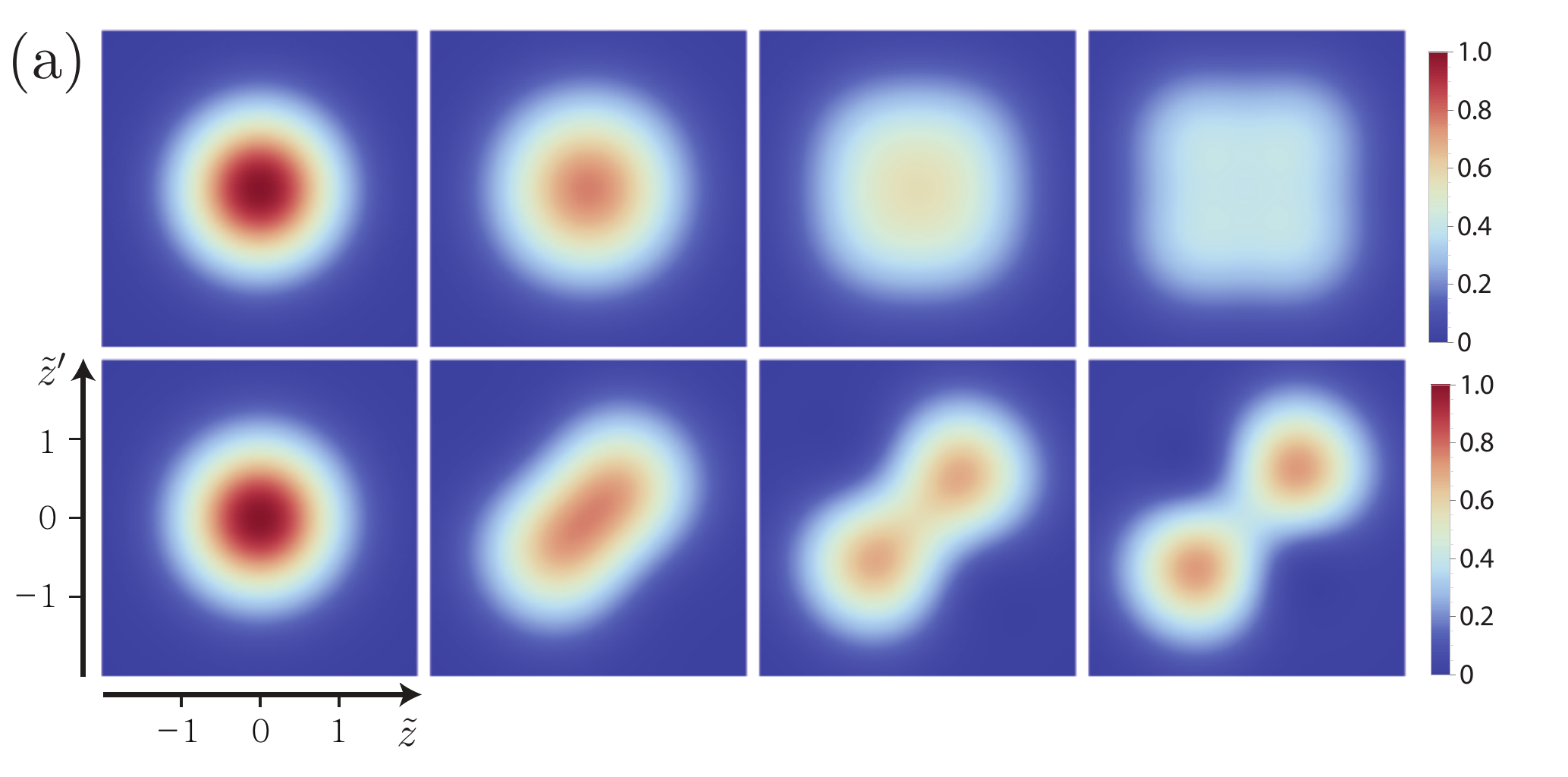}
\includegraphics[width=\columnwidth]{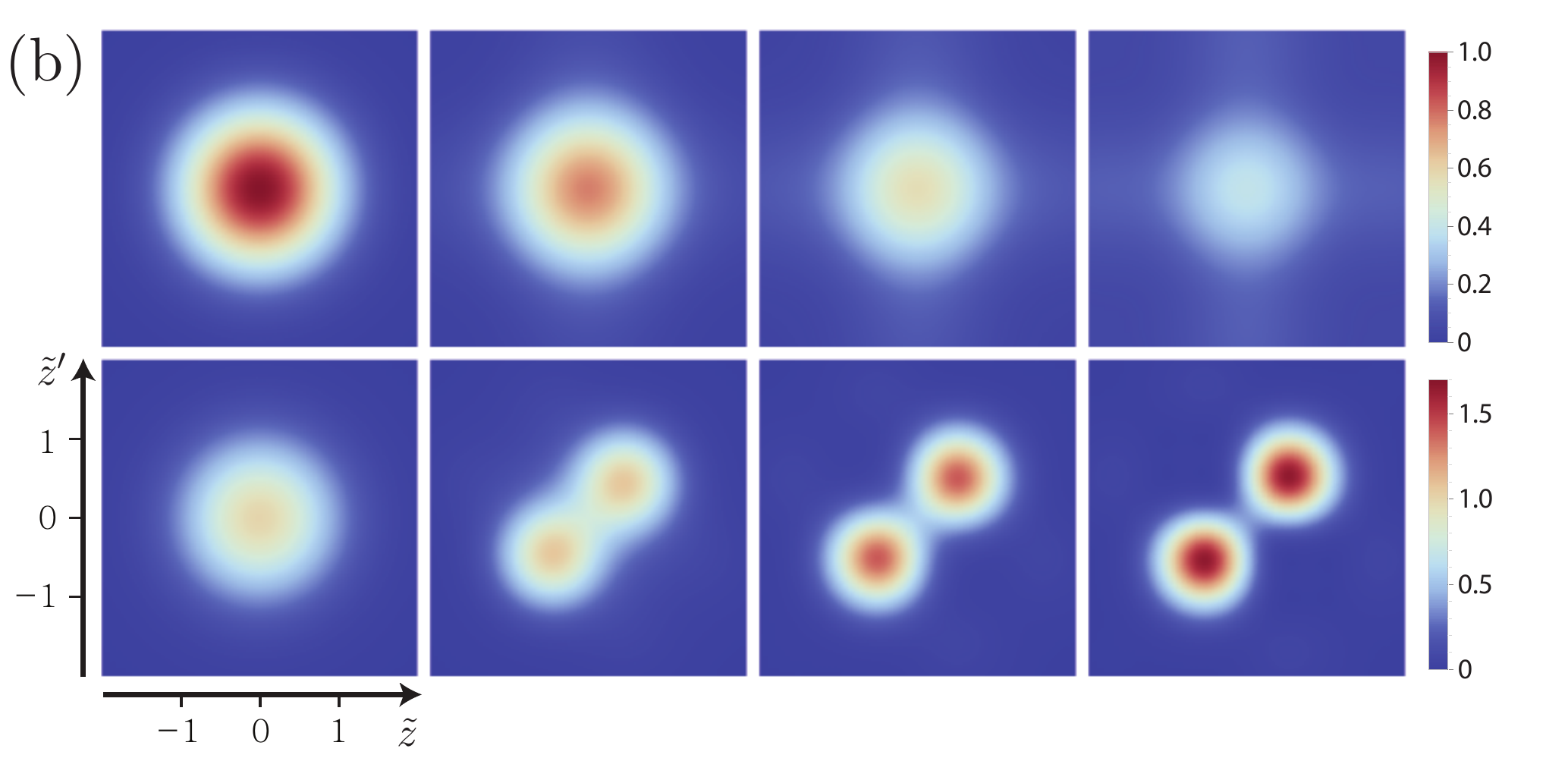}
\caption{\label{fig2} 
Temporal evolution of the density-density correlations $\rho_2(z,z',t)$ 
for $d\rightarrow 0$ (a) and $d=4$ (b) in Eq.\,\eqref{defpsi1}. 
The degree of fragmentation increases from left to right with values $\mathcal F=0,0.25,0.5,0.75$.
Top row of the panels is at $t=0$ and in original $z,z'$ variables, bottom row for $t\gg 1$ and 
in terms of scaling coordinates, $\tilde z={z}/{\sqrt{1+w_t^4}}$, and $\tilde z'$ correspondingly.
The unit of correlations is 
$N^2/[\pi(1+w_t^4)]$. Note the different color gradings at top and bottom in (b); for $\mathcal F=0$ the amplitude remains invariant between $t=0$ and $t\gg 1$.}
\end{figure}
\end{center}

Selecting, e.g., two opposite points $z=-z'$, 
{for $t\gg 1$}, we obtain the correlation ratio 
\begin{equation} \label{conc2}
\begin{split}
\frac{\rho_2(z,-z,t)}{\rho_2(z,z,t)}
=\frac{(|\tilde{\psi}_0(z,t)|^2{N}_0-|\tilde{\psi}_1(z,t)|^2{N}_1)^2}{\rho^2(z,t) 
+4{N}_0{N}_1|\tilde{\psi}_0(z,t)|^2|\tilde{\psi}_1(z,t)|^2}.
\end{split}
\end{equation} 
According to the above formula, the approximately vanishing value of $\rho_2(z,-z,t)/\rho_2(z,z,t)$ for large 
degree of fragmentation 
$\mathcal F$, visible in Fig.\,\ref{fig2}, is related to  comparable initial curvature radii of modes with given 
parity, 
i.e., to comparable dominant Fourier components. 
Note that ${\rho_2(z,-z,t)}/{\rho_2(z,z,t)}=1\;\forall\, t$ when $\mathcal F=0$, i.e., $N_0=N$.
 
We stress that when the pair coherence   $\big<\hat{a}^{\dagger}_{0}\hat{a}^{\dagger}_{0}\hat{a}_{1}\hat{a}_{1}\big>+\,{\rm h.c.}$ [cf.\,\,last term in Eq.\,\eqref{TPDMcont}] were set positive, 
the ratio in \eqref{conc2} becomes unity.
The corresponding large difference in the ratio of off-diagonal to diagonal density-density correlations 
thus allows for the confirmation of the negative sign of the macroscopic 
pair-coherence 
$\propto \ord(N^2)$. 



We make our discussion explicit by assuming the following initial orbitals set.
The harmonic oscillator ground state is used for the lower single-particle state, 
$\psi_0(z)=\pi^{-1/4}\exp\left[-{z^2}/{2}\right]$ \cite{TF}.
For the excited (odd) state, we construct a superposition of two Gaussians of opposite sign and the same width, with symmetrically placed 
centers a distance $d$ apart. This leads to 
\bea
\psi_1(z)= \frac{1}{\pi^{1/4}}\frac{\sinh\left({zd}/{2}\right)\exp\left[-{z^2}/{2}\right]}{\exp\left[d^2/16\right]\sqrt{\sinh\left(d^2/8\right)}}. \label{defpsi1}
\ea 
Varying $d$, this choice serves to illustrate the influence of the overlap {of the moduli} 
$|\psi_{0,1}(z)|$ on the correlations. For $d\rightarrow 0$ we obtain simply the first excited harmonic oscillator state, $\psi_1(z)\rightarrow \pi^{-1/4} \sqrt 2 z \exp\left[-{z^2}/{2}\right]$, 
for $d\gg 1$ the outer peaks are located where the central Gaussian {$\psi_0(z)$}
has essentially zero weight, cf. Fig.\,\ref{fig2} top. 


The hallmark of single-trap condensate fragmentation then becomes apparent upon increasing
the degree of fragmentation. As seen from Fig.\,\ref{fig2}, 
$\rho_2(z,-\alpha z,t)/\rho_2(z,z,t)$ significantly decreases in the long-time limit for 
any oppositely located points in the cloud, i.e. $z'=-\alpha z$ with $\alpha >0$. 
The robust nature of the proposed indicator is shown by decreasing the orbital overlap significantly; 
for $d=4$ in Eq.\eqref{defpsi1}, see Fig.\,\ref{fig2}\,(b), the result remains 
similar. Note that the density itself satisfies scaling invariance upon expansion of the cloud.
The density-density correlation signal thus obtained is strikingly different from that for a double well, where it exhibits Hanbury Brown-Twiss oscillations 
for $z'=-z$ and a central peak instead of the central depression seen in Fig.\,\ref{fig2}   \cite{supplement,Haroche}.  

\vspace*{-2em}

{\em Description 
with Fock-Conjugate Phase States.} 
The above results 
can be rephrased in terms of a phase state representation of fragmented condensates \cite{supplement}.
Phase states furnish the most natural tool to transparently describe coherence properties, cf., e.g.,  \cite{AndrewsKetterle,Yoo,Hadzibabic,Ashhab,Paraoanu}, and 
{will} serve to elucidate that the 
robustness of the presently discussed fragmented many-body states stems from their being conjugate to fragmented states which are (superpositions of) sharp peaks in Fock space. 

We prove in what follows that the macroscopically occupied modes of the fragmented state
correspond to sharp peaks in the distribution function corresponding to the weights
of phase states \cite{Claude}.
We define the {\em phase state representation} of 
$|\Psi\rangle$ as the integral expression 
\begin{equation}
\ket{\Psi} 
=\int^{2\pi}_0\frac{d\phi}{2\pi}C_{\phi,l_0}\ket{\phi,N,l_0} ,
\end{equation}
where $C_{\phi,l_0}=\sum_lC_l {\mathcal N}_{N,l_0;l}\,e^{-il\phi}$ with
the normalization factor ${\mathcal N}_{N,l_0;l}=\sqrt{\frac{(N-l)!l!}{N!}\frac{N^N}{(N-l_0\,)^{N-l}\,l_0^{l}}}$. The
basis vectors $\ket{\phi,N,l}=\frac{(\hat{\psi}^{\dagger}_{\phi,N,l})^N}{\sqrt{N!}}\ket{0}$ 
are created by the $l$ dependent superposition operators
\begin{equation}
\hat{\psi}^{\dagger}_{\phi,N,l}\equiv\frac{\sqrt{N-l\,}\hat{a}^{\dagger}_0+\sqrt{l\,}e^{i\phi}\hat{a}^{\dagger}_1}{\sqrt{N}}. 
\label{defOP}
\end{equation}
The phase state formulation enables us to rewrite any expectation value of an operator $\hat O$ 
in a given many-body state, to a very good approximation \cite{supplement},  
as an integral over diagonal matrix elements 
\begin{equation}
\big<\hat{O}\big> 
\simeq \int^{2\pi}_0\frac{d\phi}{2\pi}|C_{\phi}|^2\bra{\phi,N,l_0}\hat{O}\ket{\phi,N,l_0} ,
\label{phstateEV}
\end{equation}
where the amplitudes 
$C_{\phi}=\sum_lC_l e^{-il\phi}$ 
are 
the discrete Fourier transforms of the Fock space amplitudes $C_l$.

Calculating $C_{\phi}$ from the $C_l$ distribution 
of stably fragmented two-mode many-body states, 
one can show that the latter are accurately represented by    
two sharp peaks of the modulus (in the limit $N\rightarrow \infty$)   
\cite{supplement,Pi} 
\bea
|C_{\pi/2}| = |C_{3\pi/2}|
=\frac1{\sqrt2} .
\label{coeff}
\ea
This simple representation of the {\em many-body} fragmented state  in terms of two distribution peaks of phase difference $\pi$ essentially stems from the property $\textrm{sgn}(C_lC_{l+2})=-1$. 
The widths of the peaks in phase and Fock space satisfy the conjugation relation $\Delta C_{\phi} \sim (\Delta C_l)^{-1}$ ($\propto 1/\sqrt{N}$ for the Gaussian $|C_l|$ distribution), so that $\Delta C_{\phi}\rightarrow 0$ for $N\rightarrow\infty$. 
Fragmented {two-mode} condensates with {quasicontinuous} $C_l$
distributions 
hence correspond to superpositions of macroscopic states with a phase difference of $\pi$, and the two macroscopically occupied modes of the quantum gas are 
{\em globally} exactly out of phase with each other. 
This property is in sharp contrast with {double-well}  
fragmented condensates, where all values of the phase $\phi$ are equally likely ($|C_\phi|= {\rm constant}$) \cite{supplement}. 
{Macroscopically fragmented condensates are also distinct from so-called 
{\em quasicondensates} \cite{Petrov} occurring above a temperature $\propto N\omega^2/\mu$, where $\omega$ and $\mu$ are longitudinal trapping frequency and chemical potential, respectively,
which possess strongly fluctuating phases.}

The phase state formalism facilitates an interpretation of the strong suppression of $\rho_2(z,z')$ along $z=-z'$ in Fig.\,\ref{fig2} as follows. For simplicity of the following argument and notational brevity, we put ${N}_0={N}_1$ ($\mathcal{F}=1$, $l_0=N/2$), and set $\psi_1(z)$ to be the first excited harmonic oscillator state ($d\rightarrow 0$). Each of the Hilbert space vectors $\ket{\pi/2,N,\frac N2}$ and $\ket{3\pi/2,N,\frac N2}$ is a coherent state, according to the definition in Eq.\,\eqref{defOP}, for the orbitals $\psi_0(z)+i\psi_1(z)$ and $\psi_0(z)-i\psi_1(z)$, respectively, omitting the 
normalizing $1/\sqrt 2$.
After TOF ($t\gg 1$), the orbitals transform into  $\tilde{\psi}_0(\tilde{z},t)+i\tilde{\psi}_1(\tilde{z},t)$,~$\tilde{\psi}_0(\tilde{z},t)-i\tilde{\psi}_1(\tilde{z},t)$, where the scaling coordinate $\tilde z={z}/{\sqrt{1+w_t^4}}$, and 
up to an irrelevant common phase factor.
Again, $\tilde{\psi}_0(\tilde{z},t)$ is a Gaussian and now $i\tilde{\psi}_1(\tilde{z},t)$ is the first excited harmonic oscillator state. Thus $\tilde{\psi}_0(\tilde{z},t)\pm i\tilde{\psi}_1(\tilde{z},t)$ have most weight at positive and negative $z$ for upper and lower signs, respectively. From Eq. \eqref{phstateEV}, $\big<\hat{O}\big>= \frac{1}{2}\bra{\pi/2,N,\frac N2}\hat{O}\ket{\pi/2,N,\frac N2}+\frac{1}{2}\bra{3\pi/2,N,\frac N2}\hat{O}\ket{3\pi/2,N,\frac N2}$, which  decomposes into a sum of correlation functions calculated with respect to the two coherent states. Since, generally,  
$\rho_2(z,z')\simeq \rho(z)\rho(z')$ for coherent states up to $\mathcal{O}(1/N)$ terms, the resulting 
correlations will correspondingly be concentrated in the region $z,z'>0$ due to $\ket{\pi/2,N,0}$ and in the $z,z'<0$ region due to $\ket{3\pi/2,N,0}$,  but will almost vanish for $z>0,z'<0$ and $z<0,z'>0$. 
A similar argument can be carried out for $N_0\neq N_1$ and $d$ finite, so that we obtain complete
agreement with 
Fig.\,\ref{fig2}.
By the same argument, it can be shown that an absorption image of the density alone
will not allow for the unique inference that the single-trap condensate has fragmented. 

{\em Conclusion {and Outlook}.}
We have proposed an 
experimental tool using standard density-density correlation analysis to verify whether an ultracold, strongly interacting gas 
of bosons in a single trap is a 
fragmented condensate. 
The spatiotemporal behavior of density-density correlations changes dramatically with the sign and magnitude of pair-correlations between the modes. 
{Single-trap condensate fragmentation is therefore 
a {\em genuine} many-body phenomenon, in that it necessitates the observation of second-order correlations. 
By contrast, for multiple-well 
fragmentation, structure factor measurements, and hence first-order correlations, suffice to detect fragmentation: The externally imposed spatial separation 
of the fragments already entails the direct observability of vanishing off-diagonal long-range order.} 

The predicted decrease of the ratio of off-diagonal 
to diagonal density-density correlations with time should be measurable even for relatively small degrees of fragmentation $\mathcal F$. We anticipate that values of $\mathcal F$ down to the level of about 10\,\%--20\,\% should be measurable with current experimental precision.


For future work, we envisage investigating the full counting statistics of fragmented condensates.
By their very nature, there is no inverse mapping of correlation functions to a unique many-body state. 
While correlation functions can reliably measure {\em global} features of the many-body state like the degree of fragmentation, they cannot reveal local features in the Fock space distributions,  
because they integrate 
over such distributions. A single-shot analysis 
might supply 
a one-to-one mapping of the many-body state to measured quantities going beyond the predominantly Fock-state-based analyses  existing so far \cite{Shelankov}. 
Finally, 
many-body condensate fragmentation into a finite number of macroscopic pieces potentially increases the matter wave bunching 
towards the Hanbury Brown-Twiss value for a thermal cloud of bosons 
\cite{Schellekens}.

This research was supported by the NRF Korea, Grant Nos. 
2011-0029541 
and 
2014R1A2A2A01006535.


\pagebreak

\begin{widetext} 
\section{supplemental material}

\subsection{Density-density correlations for double-well fragmentation}
To contrast our result for density-density correlations in a single trap
with the well-known result for a double well \cite{Haroche}, for completeness and self-containedness
of the discussion we briefly elaborate below on the latter.

A fragmented double-well configuration describe independent condensates, i.e. simple Fock states of particle number $N_{\rm L}$ and $N_{\rm R}$, respectively. The orbitals $\psi_{\rm L} (z)$ and $\psi_{\rm R}(z)$ centers 
are displaced relative to each other by a distance $d$ due to a repulsive barrier. 
The correlation functions are given by 
\begin{equation}
\begin{split}
\big{<}\hat{\psi}^{\dagger}(z)\hat{\psi}(z)\big{>}=&N_{\rm L}|\psi_{\rm L}(z)|^2+N_{\rm R}|\psi_{\rm R}(z)|^2\\
\big{<}\hat{\psi}^{\dagger}(z)\hat{\psi}^{\dagger}(z')\hat{\psi}(z')\hat{\psi}(z)\big{>}\simeq& \left(N_{\rm L}|\psi_{\rm L}(z)|^2+N_{\rm R}|\psi_{\rm R}(z)|^2\right)\left(N_{\rm L}|\psi_{\rm L}(z')|^2+N_{\rm R}|\psi_{\rm R}(z')|^2\right)\\
&+2N_{\rm L}N_{\rm R}\Re\left[\psi^*_{\rm L}(z)\psi_{\rm L}(z')\psi^*_{\rm R}(z')\psi_{\rm R}(z)\right].\\
\end{split}\label{DWcorr}
\end{equation} 
Here, $\psi_{\rm L}(z)$ and $\psi_{\rm R}(z)$ are chosen to be two Gaussians of width $w=\sqrt{1/\omega}\equiv 1$\, each centered at $z=-d/2$, $z=d/2$ and defined as follows \cite{nonortho}
\begin{equation}
\psi_{\rm L}(z)=\frac{1}{\pi^{1/4} }\,\exp\left[-\frac{\left(z-\frac{d}{2}\right)^2}{2}\right],\quad \psi_{\rm R}(z)=\frac{1}{\pi^{1/4} }\,\exp\left[-\frac{ \left(z+\frac{d}{2}\right)^2}{2}\right].
\end{equation} 
Applying the noninteracting propagator to the initial orbitals as in Eq. \eqref{freepropagator} of the main text, the time evolution of each Gaussian under TOF can be described by $e^{i\phi(z+d/2,t)}\tilde{\psi}(z+d/2,t)$, $e^{i\phi(z-d/2,t)}\tilde{\psi}(z-d/2,t)$ where  $\tilde{\psi}(z,t)$ and $\phi(z,t)$ are
\begin{equation}
\tilde{\psi}(z,t)=\frac{1}{\pi^{1/4} (1+w_t^4)^{1/4}}\,\exp\left[-\frac{\tilde{z}^2}{2}\right], 
\qquad \tilde{z}=\frac{z}{\sqrt{1+w_t^4}},\qquad 
\phi(z,t)=\frac{1}{2t}\frac{t^2}{1+t^2}z^2-\frac{3\pi}{4}.
\end{equation} 
For $\big<\hat{\psi}^{\dagger}(z,t)\hat{\psi}(z,t)\big>$, this leads to
\begin{equation}
\big<\hat{\psi}^{\dagger}(z,t)\hat{\psi}(z,t)\big>=N_{\rm L}|\tilde{\psi}(z-d/2),t|^2+N_{\rm R}|\tilde{\psi}(z+d/2),t|^2.
\end{equation} 
The expected average of density in many experimental runs is just a Gaussian profile with normalization given by the total number of particles $N_{\rm L}+N_{\rm R}$.

On the other hand, the density-density correlation function furnishes nontrivial features, in form of 
Hanbury Brown-Twiss (HBT) correlations, for which the above defined phase factor $\phi(z,t)$ plays the major role \cite{Haroche}
\begin{equation} \label{ddcorrphase1}
\begin{split}
\big{<}\hat{\psi}^{\dagger}(z,t)&\hat{\psi}^{\dagger}(z',t)\hat{\psi}(z',t)\hat{\psi}(z,t)\big{>}\\
\simeq
&\left(N_{\rm L}|\tilde{\psi}(z-d/2,t)|^2+N_{\rm R}|\tilde{\psi}(z+d/2,t)|^2\right)\left(N_{\rm L}|\tilde{\psi}(z'-d/2,t)|^2+N_{\rm R}|\tilde{\psi}(z'+d/2,t)|^2\right)\\
&+2N_{\rm L}N_{\rm R} \left[\tilde{\psi}(z-d/2,t)\tilde{\psi}(z'-d/2,t)\tilde{\psi}(z'+d/2,t)\tilde{\psi}(z+d/2,t)\right] \cos\left(\frac{1}{t}\frac{t^2}{1+t^2}(z-z')d\right).
\end{split}
\end{equation} 
For $t\gg 1$, the HBT term becomes
\begin{equation}\label{BECBECHBTTOF}
2N_{\rm L}N_{\rm R} \left[\tilde{\psi}(z-d/2,t)\tilde{\psi}(z'-d/2,t)\tilde{\psi}(z'+d/2,t)\tilde{\psi}(z+d/2,t)\right] \cos\left[d(\tilde{z}-\tilde{z}')\right].
\end{equation} 
The term in square brackets reduces to\, $\simeq|\tilde{\psi}(z)|^2|\tilde{\psi}(z')|^2$ as $\sqrt{1+w_t^4}\simeq t\gg d$. 
Looking at the cosine part, $(\tilde{z}-\tilde{z'})$ is scale-invariant, thus the initial $d$  determines the correlation oscillation features in the long time limit. For $t\gg d$ and $t\gg 1$, we then have approximately
\begin{equation}
\big{<}\hat{\psi}^{\dagger}(z,t)\hat{\psi}^{\dagger}(z',t)\hat{\psi}(z',t)\hat{\psi}(z,t)\big{>}\simeq 
|\tilde{\psi}(z,t)|^2|\tilde{\psi}(z',t)|^2\left[N^2_{\rm L}+N^2_{\rm R}+2N_{\rm L}N_{\rm R}\left(1+\cos\left[d(\tilde{z}-\tilde{z}')\right]\right)\right].
\end{equation}
In Fig.\,\ref{figsuppl}, we plot the correlations before and after TOF for separations
$d=4,6,8$, illustrating the development of fringes in the 
off-diagonal direction $z'=-z$. One should compare these plots with those shown in 
Fig.\,\ref{fig2} of the main text: 
{In a single trap fragmented state, there are no such density-density-correlation interference fringes 
to be detected, also cf. the discussion at the end of the next section.}
 
These considerations can be extended to, e.g., triple wells, which show qualitatively very similar correlation features. The basic differences in the correlation signal between single-trap and multi-well configurations 
are therefore 
not related to the number of maxima
in the total density.
\begin{center}
\begin{figure}[t]
\includegraphics[width=0.6\columnwidth]{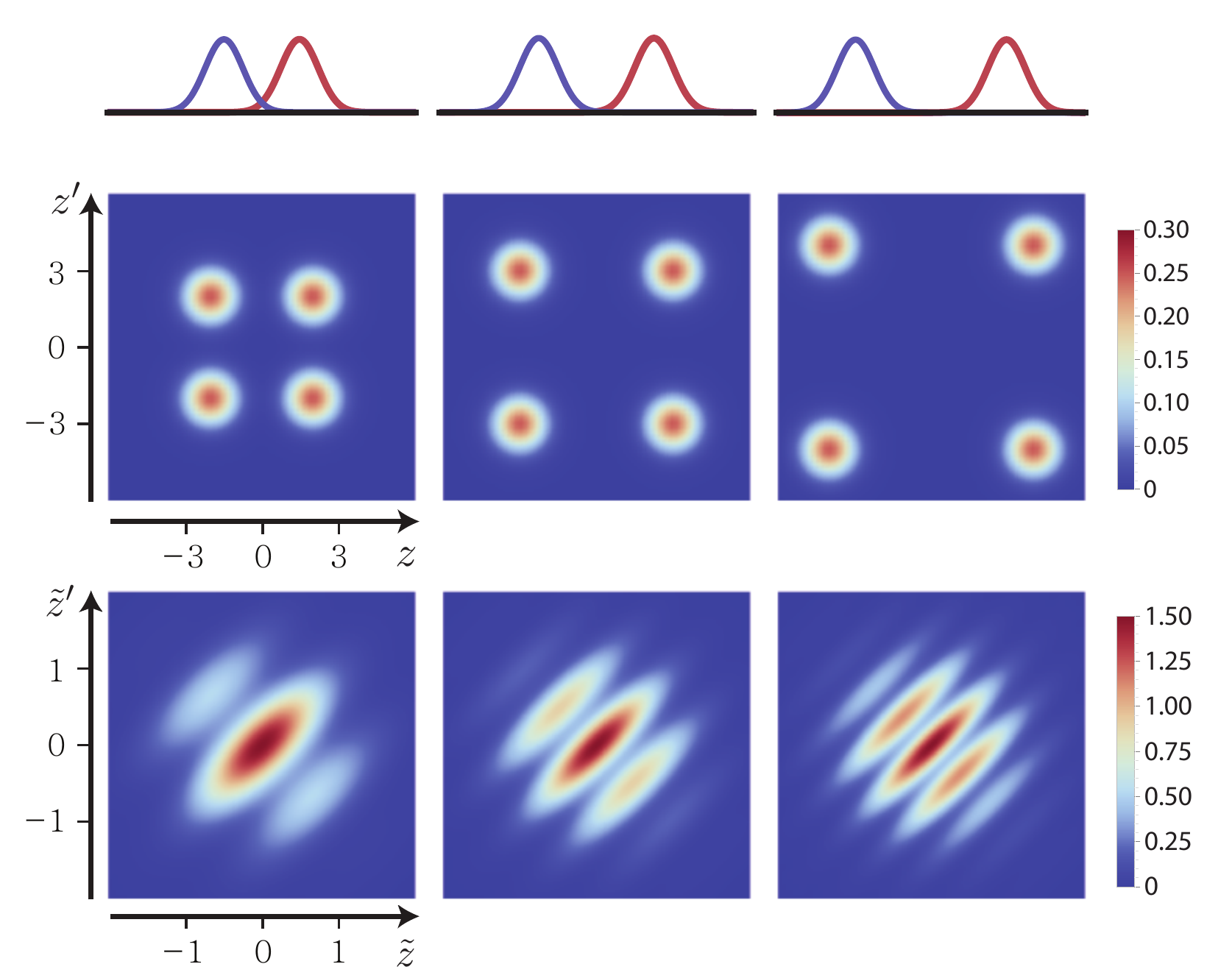}
\caption{\label{figsuppl} Density-density correlations of a symmetric double-well fragmented state ($N_{\rm L}=N_{\rm R}$)  
before (top) and after (bottom) TOF. The correlation unit is ${N^2}/[\pi(1+w_t^4)]$. } 
\vspace*{-1em}
\end{figure}
\end{center}

\subsection{Phase state formalism}

In the literature, cf., e.g. \cite{Claude}, 
the phase state formalism used in our paper to illustrate the coherence properties of stably
fragmented states  is commonly applied to very specific many-body states, 
in particular, (superpositions of) single Fock states or coherent states. In addition, a proper analysis of its domain of applicability is generally missing. We therefore provide in this supplement such an analysis of the validity of the phase state formalism for general, {quasicontinuous} Fock-state-amplitude 
many-body states, in particular with respect to the accurate evaluation of the experimental observables, i.e., correlation functions.

We begin our discussion with the known example of a single Fock state $\ket{N-l,l}$. 
The latter can be written as a linear combination of phase states $\ket{\phi,N}$ as follows \cite{Claude} 
\begin{equation}
\ket{N-l,l}=\frac{(\hat{a}^{\dagger}_0)^{N-l}(\hat{a}^{\dagger}_1)^{l}}{\sqrt{(N-l)!l!}}\ket{0}=\int^{2\pi}_0\frac{d\phi}{2\pi}\,\sqrt{\frac{(N-l)!l!}{N!}2^N}\,e^{-il\phi}\ket{\phi,N} 
\end{equation}
where the phase state $\ket{\phi,N}$ is defined as
\begin{equation}
\begin{split}
&\ket{\phi,N}=\frac{(\hat{\psi}^{\dagger}_{\phi})^N}{\sqrt{N!}}\ket{0},~\hat{\psi}^{\dagger}_{\phi}=\frac{\hat{a}^{\dagger}_0+e^{i\phi}\hat{a}^{\dagger}_1}{\sqrt{2}},\quad\hat{\psi}(\bm{r})\ket{\phi,N}=\sqrt{N}\psi_{\phi}({\bm r})\ket{\phi,N-1},~\psi_{\phi}({\bm r})\equiv\left[\hat{\psi}({\bm r}),\hat{\psi}^{\dagger}_{\phi}\right]\\
&\braket{\phi',N}{\phi,N}=\exp\left[i\frac{N(\phi-\phi')}{2}\right]\left(\cos\left[\frac{\phi-\phi'}{2}\right]\right)^N .
\end{split}\label{defNphi}
\end{equation}
In terms of $\ket{\phi,N}$, the expectation value of the density, $\hat{\rho}({\bm r})=\hat{\psi}^{\dagger}({\bm r})\hat{\psi}({\bm r})$ 
can be written as a double integration over two phase angles $\phi$ and $\phi'$  
\begin{equation}
\bra{N-l,l}\hat{\rho}({\bm r})\ket{N-l,l}=\frac{(N-l)!l!}{N!}2^N\int^{2\pi}_0\frac{d\phi'}{2\pi}\int^{2\pi}_0\frac{d\phi}{2\pi}\,e^{i(N-2l-1)\Delta\phi}(\cos\Delta\phi)^{N-1}\psi^*_{\phi'}({\bm r})\psi_{\phi}({\bm r}), 
\end{equation}
where $\Delta\phi=(\phi-\phi')/2$. 

In the large $N$ limit, the $N$-th power of $\cos\Delta\phi$ is approximately $e^{-N(\Delta\phi)^2/2}$, with a value $\mathcal{O}(1)$ within the range $|\Delta\phi|<\pi/\sqrt{N}$. 
Thus we can safely reduce the double integral into an integral over the single phase $\phi$ by putting $\phi'\simeq\phi$ and approximate the exponential factor by unity provided $N-2l\ll\sqrt{N}$. For the case of the evenly distributed single Fock state,  $l=N/2$ 
the following approximate equality is therefore obtained, cf.\,\cite{Pethick} chapter 13,  
\begin{equation}
\bra{N/2,N/2}\hat{\rho}({\bm r})\ket{N/2,N/2}\simeq\int^{2\pi}_0\frac{d\phi}{2\pi} \bra{\phi,N}\hat{\rho}({\bm r})\ket{\phi,N}.
\label{NNexpect}
\end{equation}
Thus a Fock state $\ket{N/2,N/2}$ can be interpreted as an ensemble of all phase (coherent) states $\ket{\phi,N}$ 
with equal probability \cite{Yoo}.
This result is applicable not only for $\hat{\rho}({\bm r})$ but also for any $n$-body operator $\hat{O}_n$ where $n\ll N$ when $N\rightarrow\infty$ \cite{Mueller}. That any $\phi$ will be measured with equal probability was experimentally shown with interference fringes resulting from the TOF overlap of two initially independent BECs. The offset of fringes was different for each experimental run \cite{AndrewsKetterle}; this was later on confirmed for the interference of thirty condensates released from optical lattice wells \cite{Hadzibabic}. Theoretically, the concept of phase states was previously applied 
to time of flight experiments for weakly depleted condensates \cite{Paraoanu},  
and for the measurement theory of many-body states (counting statistics) in \cite{Shelankov}.

For a general $\ket{N-l,l}$, when $l\neq N/2$, we redefine $\hat{\psi}^{\dagger}_{\phi}$ and correspondingly  $\ket{\phi,N}$ and $\psi_{\phi}({\bm r})$ as follows \cite{Shelankov, Claude} 
\begin{equation}
\hat{\psi}^{\dagger}_{\phi,N,l}=\frac{\sqrt{N-l}\hat{a}^{\dagger}_0+e^{i\phi}\sqrt{l}\hat{a}^{\dagger}_1}{\sqrt{N}},~~\ket{\phi,N,l}=\frac{(\hat{\psi}^{\dagger}_{\phi,N,l})^N}{\sqrt{N!}}\ket{0},~~\psi_{\phi,N,l}({\bm r})=\left[\hat{\psi}({\bm r}),\hat{\psi}^{\dagger}_{\phi,N,l}\right].
\end{equation}
We now calculate the expectation value  of an arbitrary normal-ordered $n$-body operator 
\begin{equation}
\mathopen{:}\hat{O}_n\mathclose{:}=\mathopen{:}\, \prod_{i=1}^n\hat{\psi}^{\dagger}({\bm r}_i)\hat{\psi}({\bm r}_i)\, \mathclose{:}. 
\label{:On:}
\end{equation}
We are going to show that the expectation value of \eqref{:On:} can be computed in the form of \eqref{NNexpect}. 
This, then, allows us to understand the phase state $\ket{\phi,N,l}$ as an approximate eigenstate of the operator $\hat O_n$. For a simple $n=2$ example, 
$\bra{\phi,N,l} \hat \rho ({\bm r}_1) \hat \rho ({\bm r}_2 )\ket{\phi,N,l} 
\simeq \bra{\phi,N,l}:\hat \rho ({\bm r}_1) \hat \rho ({\bm r}_2 ):\ket{\phi,N,l}
\simeq \bra{\phi,N,l}\hat \rho ({\bm r}_1)\ket{\phi,N,l} \bra{\phi,N,l}\hat \rho ({\bm r}_2 )\ket{\phi,N,l}$.  

We first obtain that 
\begin{equation}
\begin{split}
\bra{N-l,l}\mathopen{:}\hat{O}_n\mathclose{:}\ket{N-l,l}  &\simeq \int^{2\pi}_0\frac{d\phi'}{2\pi}\int^{2\pi}_0\frac{d\phi}{2\pi}\,\sqrt{\frac{2\pi(N-l)l}{N}}
\left(\sum_{j=0}^{N-n}\frac{N!}{(N-n-j)!j!}\frac{(N-l)^{N-n-j}\,l^j}{N^N}\,e^{-i(l-j)(\phi-\phi')}\right) \\
&\hspace*{-4em}\times\left[\prod_{i=1}^n\left\{
(N-l)|\psi_0({\bm r}_i)|^2+e^{i(\phi-\phi')}l|\psi_1({\bm r}_i)|^2 +\sqrt{(N-l)l}\left(e^{i\phi}\psi^*_0({\bm r}_i)\psi_1({\bm r}_i)+e^{-i\phi'}\psi_0({\bm r}_i)\psi^*_1({\bm r}_i) 
\right)\right\}\right].
\end{split}
\end{equation}
Here, the summation over $j$ stems from the binomial expansion of 
$\left[\hat{\psi}_{\phi',N,l},\hat{\psi}^{\dagger}_{\phi,N,l}\right]^{N-n}$, corresponding to $(\cos\Delta\phi)^{N-n}$ for $l=N/2$. Integrating over $\phi$ for the simple $n=1$ case, we get
\begin{equation}
\begin{split}
&\int^{2\pi}_0\frac{d\phi}{2\pi}\left[(N-l)|\psi_0({\bm r}_1)|^2+l|\psi_1({\bm r}_1)|^2+\sqrt{(N-l)l}\left(e^{i\phi}\psi^*_0({\bm r}_1)\psi_1({\bm r}_1)+e^{-i\phi}\psi_0({\bm r}_1)\psi^*_1({\bm r}_1)\right)\right]\\
=&\int^{2\pi}_0\frac{d\phi}{2\pi}N\psi^*_{\phi,N,l}({\bm r}_1)\psi_{\phi,N,l}({\bm r}_1)
=\int^{2\pi}_0\frac{d\phi}{2\pi}\bra{\phi,N,l}\hat{\rho}({\bm r}_1)\ket{\phi,N,l}, 
\end{split}
\end{equation}
which is the desired result. Evaluating analogously the expectation value for general $n$, we obtain 
\begin{equation} \label{ayo}
\begin{split}
\bra{N-l,l}\mathopen{:}\hat{O}_n\mathclose{:}\ket{N-l,l}\simeq 
&\int^{2\pi}_0\frac{d\phi}{2\pi}\prod^n_{i=1}(N-n+1)\psi^*_{\phi,N,l}({\bm r}_i)\psi_{\phi,N,l}({\bm r}_i)\\
=&\int^{2\pi}_0\frac{d\phi}{2\pi}\bra{\phi,N,l}\hat{O}_n\ket{\phi,N,l}, 
\end{split}
\end{equation}
where the approximate equality in the first line holds as long as for $0\leq b \leq n$ \cite{remark}
\begin{equation} \label{ayo1}
\begin{split}
&\left(\frac{(N-l)\cdots(N-l-(n-b)+1)}{(N-l)^{n-b}}\right)\,\left(\frac{l\cdots(l-b+1)}{l^{b}}\right) \\
&=\left(\prod^{n-b}_{j=1}\left(1-\frac{j-1}{N-l}\right)\right)\,\left(\prod^{b}_{k=1}\left(1-\frac{k-1}{l}\right)\right)\simeq\left(\prod^{n}_{j=1}\left(1-\frac{j-1}{N}\right)\right). 
\end{split}
\end{equation}
We note that the above proof is valid for $\mathopen{:}\hat{O}_n\mathclose{:} =\mathopen{:}~\prod^n_{i=1}\hat{\psi}^{\dagger}({\bm r}_i\,')\hat{\psi}({\bm r}_i)~\mathclose{:}$, where ${\bm r}_i\,'\neq{\bm r}_i$, 
and thus holds for any $n$-body operator.

For a general two-mode many body state $\ket{\Psi}=\sum_lC_l\ket{N-l,l}$, we do not have an exact number state.
Therefore, we have to carefully  select the appropriate $l$ value to evaluate correlation functions in some given order.
We will now show that we are able to obtain a weighed average of \eqref{ayo} to achieve this task.  
We assume that the $C_l$ distribution is centered on one specific value $l=l_0$,
and define $l_{\rm min}<l_0$ and $l_{\rm max}>l_0$ to which the distribution extends from that central value as follows.
\begin{equation}
\sum_{l\leq l_{\rm min}}|C_l|^2\leq \epsilon ,\quad  \sum_{l\geq l_{\rm max}}|C_l|^2\leq \epsilon, \qquad \epsilon \ll 1 .
\end{equation}
By writing $\ket{\Psi}$ in terms of integrals of $\phi'$ and $\phi$ and integrate over $\phi'$, we get the following expression for the density expectation value, $n=1$
\begin{equation} \label{aayo}
\begin{split}
\big<\hat{\rho}({\bm r}_1)\big>_{\Psi}&=\int^{2\pi}_0\frac{d\phi}{2\pi}\sum_{l,l'}
C^*_{l'}C_l\frac{{\mathcal N}_{N,l_0;l}}{{\mathcal N}_{N,l_0;l'}}\,e^{-i(l-l')\phi}\bigg[(N-l')|\psi_0({\bm r}_1)|^2+l'|\psi_1({\bm r}_1)|^2\\
&+\sqrt{(N-l_0)l_0}\left(\frac{N-l'}{N-l_0\vphantom{\hat{N}}}e^{-i\phi}\psi_0({\bm r}_1)\psi^*_1({\bm r}_1)+\frac{l'}{l_0\vphantom{\hat{N}}}e^{i\phi}\psi^*_0({\bm r}_1)\psi_1({\bm r}_1)\right)\bigg],
\end{split}
\end{equation}
where ${\mathcal N}_{N,l_0;l}$ is defined as
\begin{equation}
{\mathcal N}_{N,l_0;l}=\sqrt{\frac{(N-l)!l!}{N!}\frac{N^N}{(N-l_0\,)^{N-l}\,l_0^{l}}}\simeq\sqrt{\left(\frac{N-l}{N-l_0\vphantom{\hat{N}}}\right)^{N-l}\left(\frac{l}{l_0\vphantom{\hat{N}}}\right)^{l}}\left(\frac{2\pi(N-l)l}{N}\right)^{1/4} .
\end{equation}
The expression (\ref{aayo}) looks complicated, but since only $l=l'$ and $l=l'\pm1$ give nonvanishing contributions in the integration over $\phi$, one can set ${\mathcal N}_{N,l_0;l}\simeq {\mathcal N}_{N,l_0;l'}$. Considering the sum for $l_{\rm min}<l,l'<l_{\rm max}$, 
if $|l-l_{\rm min}|, |l-l_{\rm max}|$ is much smaller than both $N-l_0$ and $l_0$, we can approximate (\ref{aayo}) as
\begin{equation} \label{aayoZ}
\begin{split}
\big<\hat{\rho}({\bm r}_1)\big>_{\Psi}&\simeq\int^{2\pi}_0\frac{d\phi}{2\pi}\sum_{l,l'} C^*_{l'}C_l\,e^{-i(l-l')\phi}\bigg[(N-l_0)|\psi_0({\bm r}_1)|^2+l_0|\psi_1({\bm r}_1)|^2\\
&+\sqrt{(N-l_0)l_0}\left(e^{-i\phi}\psi_0({\bm r}_1)\psi^*_1({\bm r}_1)+e^{i\phi}\psi^*_0({\bm r}_1)\psi_1({\bm r}_1)\right)\bigg]\\
&=\int^{2\pi}_0\frac{d\phi}{2\pi}\,|C_{\phi}|^2\bra{\phi,N,l_0}\hat{\rho}({\bm r}_1)\ket{\phi,N,l_0},
\end{split}
\end{equation}
where the {\em phase state amplitudes} are defined to be  
\begin{equation}
C_{\phi}\equiv\sum_l C_l e^{-il\phi}. \label{aayo1}
\end{equation}
We now consider a general operator $\mathopen{:}\hat{O}_n\mathclose{:}$. We perform the following approximation
\begin{equation} \label{ayo2}
\begin{split} 
\left(\frac{(N-l)\cdots(N-l-(n-b)+1)}{(N-l)^{n-b}}\right)\,\left(\frac{l\cdots(l-b+1)}{l^{b}}\right) 
\simeq&\left(\frac{(N-l)\cdots(N-l-(n-b)+1)}{(N-l_0)^{n-b}}\right)\,\left(\frac{l\cdots(l-b+1)}{l_0^{b}}\right),
\end{split}
\end{equation}
 for $l_{\rm min}<l<l_{\rm max}$. Then again every term contained in $l,l'$ summation gets a common prefactor by applying (\ref{ayo1}), thus we can concisely write $\bra{\Psi}\mathopen{:}\hat{O}_n\mathclose{:}\ket{\Psi}$ as
\begin{equation} \label{result1}
\bra{\Psi}\mathopen{:}\hat{O}_n\mathclose{:}\ket{\Psi}\simeq \int^{2\pi}_0\frac{d\phi}{2\pi}\,|C_{\phi}|^2\bra{\phi,N,l_0}\mathopen{:}\hat{O}_n\mathclose{:}\ket{\phi,N,l_0},
\end{equation}
with $C_{\phi}$ defined in (\ref{aayo1}). 

The error incurred by changing $l$ to $l_0$ in the denominator of (\ref{ayo2}) can be estimated by evaluating the 
maximum of the following four numbers 
\begin{equation}
\left|1-\left(\frac{N-l_{\rm min}}{N-l_0}\right)^n\right|,~\left|1-\left(\frac{l_{\rm min}}{l_0}\right)^n\right|,~\left|1-\left(\frac{N-l_{\rm max}}{N-l_0}\right)^n\right|,~\left|1-\left(\frac{l_{\rm max}}{l_0}\right)^n\right|,
\end{equation}
when $n$ is sufficiently small and $N$ is large enough. This proof is also valid for $\mathopen{:}~\prod^n_{i=1}\hat{\psi}^{\dagger}({\bm r}'_i)\hat{\psi}({\bm r}_i)~\mathclose{:}$, where ${\bm r}_i\,'\neq{\bm r}_i$ thus again holds for any $n$-body operator.

\begin{center}
\begin{figure}[t]
\includegraphics[width=.4\textwidth]{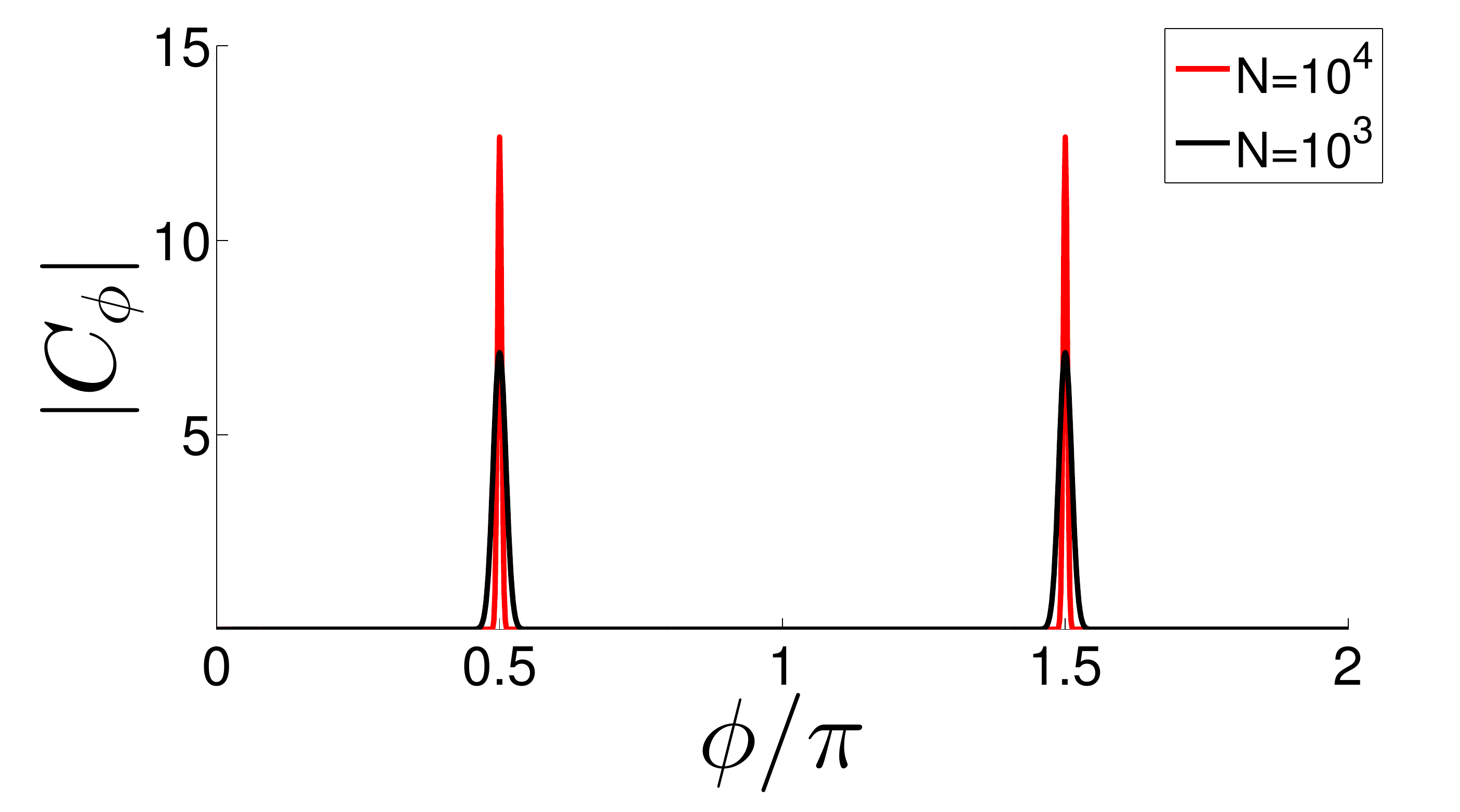}
\includegraphics[width=.4\linewidth]{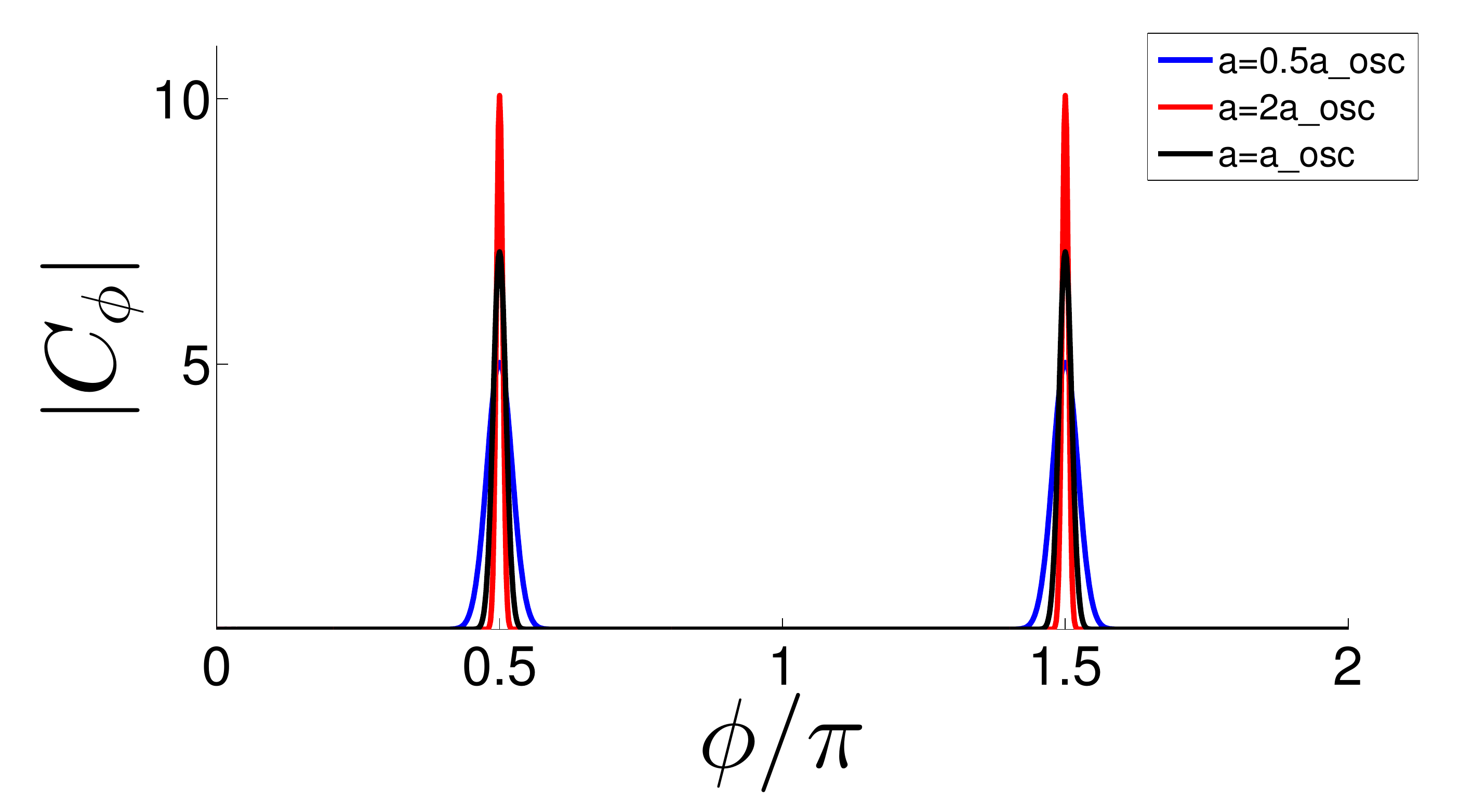} 
\caption{\label{F} Left: For maximal fragmentation $\mathcal{F}=1$ ($\mathcal{S}=0$), $N=1000$, the modulus $|C_{\phi}|$
is centered at $\pi/2$ and $3\pi/2$ and the width $\Delta |C_{\phi}|\sim \pi/\sqrt{N}\simeq0.1$. In red we show the distribution for $N=10000$, all other parameters identical; then $\Delta |C_{\phi}|\sim \pi/\sqrt{N}\simeq0.03$. Right: Variation of the width of the $|C_{\phi}|$ distribution upon increasing or decreasing the width $a$ in the Gaussian amplitude distribution \eqref{contlimit}. All other parameters identical
to plots on left.
}
\end{figure}
\end{center}

We now investigate the properties of the phase state amplitude 
$C_{\phi}$. This is a discrete Fourier transform of $C_l$; thus we expect a canonical relation between $C_{\phi}$ and $C_l$,  giving a Heisenberg indeterminacy relation of the form   
\begin{equation} \label{canon}
\Delta |C_{\phi}|\Delta |C_l| \sim 1 .
\end{equation}
As an example, we will consider the continuum approximation for the two-mode Hamiltonian discussed in \cite{Bader}. 
Then the magnitude $|C(l)|$ is a shifted Gaussian:
\begin{equation} \label{contlimit}
|C(l)|=\frac{1}{\left(\pi a^2\right)^{1/4}}\textrm{exp}\left[-\frac{\left(l-\frac{N}{2}-\mathcal{S}\right)^2}{2a^2}\right].
\end{equation} 
According to \cite{Bader}, a fragmented state has $\textrm{sgn}(C_l C_{l\pm 2})=-1$ with the ``oscillator width'' given by  $a_{\rm osc}=(2/3)^{1/4}\,\sqrt N$ \cite{Fischer}.  
Fig.\,\ref{F} left shows two particular examples for the resulting $C_{\phi}$ distribution. The degree
of fragmentation  $\mathcal{F}$ does not affect the relative heights of the peaks in the distribution $|C_{\phi}|$ \cite{relativephase}. 
In Fig.\ref{F} right we verify the expectation, based on \eqref{canon}, that the  $C_{\phi}$ distribution
becomes wider the smaller $a$ is (and thus the more narrow the $|C_l|$ distribution).

{
For a fragmented  condensate many-body state $\ket{\Psi}$ in the natural basis 
which can be expressed as a superposition of phase states, $\ket{\Psi}=\int d\phi~C_{\phi} \ket{\phi, N}$, 
the condition 
$\big<\hat{a}^{\dagger}_0\hat{a}_1\big>=0$ leads to 
\begin{equation}
\int_0^{2\pi} d\phi ~|C_{\phi}|^2e^{i\phi}=0. \label{fragcondition}  
\end{equation} 
The corresponding $C_{\phi}$ distribution for the single-trap fragmented state has two peaks, at values of $\phi$ separated by $\pi$. They are symmetrically located at $\phi=\pi/2,~3\pi/2$
for the state discussed in the main text.}

{The distribution of constant $|C_{\phi}|$ of a double-well fragmented state in the left- and right-well basis obviously also satisfies \eqref{fragcondition}. We now compare the two different types of fragmented state,
double well and single trap, by their density-density correlation function $\rho_2(z,z')$, using
their respective $C_\phi$ distributions. Let us assume that we have a many-body state which can be described by a phase state distribution satisfying \eqref{fragcondition}. For easy and direct comparison with the double well discussed in the preceding section, we write the formulas below in one spatial dimension, noting that all results can be readily generalized to arbitrary dimension.
The density $\rho(z)$ is given as $\rho(z)=N_0|\psi_0(z)|^2+N_1|\psi_1(z)|^2$ uniquely by using \eqref{result1} and \eqref{fragcondition}. Therefore, $\rho(z)$ does not reveal any details of the $C_{\phi}$ distribution. For the second-order correlations, on the other hand, we have 
\begin{equation} \label{result2}
\begin{split}
\rho_2(z,z')&=\int^{2\pi}_{0} \frac{d\phi}{2\pi}\,\,|C_{\phi}|^2\left(\rho(z)+\sqrt{N_0N_1}\left(e^{i\phi}\psi^*_0(z)\psi_1(z)+e^{-i\phi}\psi_0(z)\psi^*_1(z)\right)\right)\times\left(\vphantom{\sqrt{N_0}}z\rightarrow z'\right)\\
&=\int^{2\pi}_{0} \frac{d\phi}{2\pi}\,\,|C_{\phi}|^2\left(\rho(z)\rho(z')+2N_0N_1\Re\left[\psi^*_0(z)\psi^*_1(z')\psi_0(z')\psi_1(z)+e^{2i\phi}\psi^*_0(z)\psi^*_0(z')\psi_1(z')\psi_1(z)\right]\right), 
\end{split}
\end{equation}
where \eqref{fragcondition} is used in the second line. We now note that the integration of $|C_{\phi}|^2e^{2i\phi}$ over $\phi$ can depend on details of the $C_{\phi}$ distribution. For double-well fragmentation, $|C_{\phi}|$ is constant for all $\phi$, so that $\rho_2(z,z')$ becomes
\begin{equation}
\rho_2(z,z')=\rho(z)\rho(z')+2N_0N_1\Re\left[\psi^*_0(z)\psi^*_1(z')\psi_0(z')\psi_1(z)\right].
\end{equation}
Thus the term $\propto e^{2i\phi}$ in the second line of \eqref{result2} 
vanishes after integration, and only the HBT correlation term in Eq.\,\eqref{DWcorr} ($0\rightarrow {\rm L},1\rightarrow {\rm R}$) survives apart from the simple product of $\rho(z)$ and $\rho(z')$. 
Turning to the single-trap fragmented state, which has a $C_{\phi}$ distribution with two peaks at $\phi=\pi/2,3\pi/2$, we obtain 
\begin{equation}
\rho_2(z,z')=\rho(z)\rho(z')+2N_0N_1\Re\left[\psi^*_0(z)\psi^*_1(z')\psi_0(z')\psi_1(z)-\psi^*_0(z)\psi^*_0(z')\psi_1(z')\psi_1(z)\right].
\end{equation}
The correlation function hence acquires a term distinct from HBT, which stems from the two-peak structure of the $C_{\phi}$ distribution.
We therefore conclude that the phase-state analysis shows that a single-trap fragmented state can be distinguished from a double-well fragmented state not only due to the absence of HBT terms in the density-density correlations, but also because of the existence of an additional non-HBT correlation term.
}

\end{widetext}

\begin{thebibliography}{9999}
\bibitem{remarkTPDM} We neglect the difference between the TPDM and the density-density correlation
function which is of order $1/N$, and corresponds to quantum shot noise \cite{Gomes}.
\bibitem{expect} We expect that our results are robust against weakly broken parity, cf.\,the extensive discussion in \cite{FischerII}.
\bibitem{NPCremark} Note that the magnitude and sign of the pair coherence ensures that 
the (suitably regularized) variance $\big< \left(\hat \rho(z) -\rho(z)\right)^2 \big> \simeq \rho_2(z,z)- \rho^2(z)$  
is $\ord (N)$, implying that the trapped fragmented condensate is an (approximate) eigenstate of the density operator.
\bibitem{ballnote} {Putting the orbital width parameter $w$ to be of order $R_{\rm TF}$, the TF size of the initial cloud (suitably generalized for a fragmented condensate \cite{Fischer}), ballistic expansion begins when $t\gg \ord (R_{\rm TF}^2)$, also cf.\,Ref.\,\cite{Gomes}.}
\bibitem{TF} We have verified that the density-density correlations after TOF 
demonstrate the same qualitative behavior at late times, choosing, e.g., a TF ground state wavefunction  $\psi_{0,{\rm TF}}=(3(4-z^2)/{32})^{1/2}$, for $-2<z<2$, and zero elsewhere, instead of a Gaussian.
\bibitem{Pi} Note that the (symmetric around $\pi$) absolute location 
of the distribution maxima stems from the definition \eqref{defOP}, 
while their phase difference $\pi$ is physical.
\end{thebibliography}
\begin{thebibliography}{99}
\bibitem[S1]{nonortho} We neglect the slight nonorthogonality (i.e., due to finite overlap) of the Gaussians 
for large $d$.
\bibitem[S2]{remark} Performing the calculation for any $n$, after integrating over $\phi'$ one obtains an expression for  $\big<\mathopen{:}\hat{O}_n\mathclose{:}\big>$ in terms of an integration over $\phi$.
Assuming (\ref{ayo1}), separating off the prefactor $\prod^n_{i=1}(N-n+1)$ enables us to write the result in the form of (\ref{ayo}). When both $N-l$ and $l$ are $\mathcal{O}(N)$, and $N\rightarrow \infty$, $n\ll N$, it can be shown 
for any $n$ that in fact it is sufficiently accurate to use (\ref{ayo1}). For example, when $N=1000$, $l=N-l=500$, for $n=5$ the difference between left-hand side and right-hand side of (\ref{ayo1}) is about $1.5\%$, for $l=100$ 
the difference is $5\%$ when $n=3$ and $13\%$ when $n=5$. 
Therefore it can be concluded that for experimentally directly accessible order of correlation functions (up to third order, $n=3$, currently), Eq.\,\eqref{ayo1} is reliable even for relatively small, mesoscopic $N=1000$. 
\bibitem[S3]{relativephase} This is true when the phase difference between even and odd $l$ sectors, 
$\theta\equiv \textrm{Arg}(C_{l+1}/C_{l})$, is zero, $\theta=0$, as assumed throughout our analysis. 
The degree of fragmentation becomes progressively smaller increasing
$\theta$ towards $\theta=\pi/2$, where it is of $\ord (1/N)$, that is essentially vanishes, cf.\,\cite{Lee}. The two peaks then transform into the single peak occurring for a single condensate, maintaining the normalization $|C_\frac{\pi}{2}|^2+|C_\frac{3\pi}{2}|^2=1$.
\end{thebibliography}
\end{document}